\documentclass[english,12pt]{article}
\usepackage{array}
\usepackage{graphicx}
\usepackage{amssymb}
\usepackage{amsmath}
\usepackage{multirow}
\usepackage{prettyref}
\usepackage{babel}
\usepackage{units}
\usepackage[latin1]{inputenc}
\usepackage{amsfonts}
\usepackage{amssymb}
\usepackage{babel}
\usepackage{perpage} %the perpage package
\MakePerPage{footnote} %the perpage package command
\usepackage{cite}
\def\@fmsl@sh#1#2#3{\m@th\ooalign{$\hfil#1\mkern#2/\hfil$\crcr$#1#3$}}
 \def\eq#1\en{\begin{equation}#1\end{equation}}
\def\s[#1,#2]{[#1\stackrel{\star}{,}#2]}
\def\sx[#1,#2]{[#1\stackrel{\star_{x}}{,}#2]}

\newcommand{\nc}{\newcommand}
\nc{\beq}{\begin{equation}}
\nc{\eeq}{\end{equation}}
\nc{\beqa}{\begin{eqnarray}}
\nc{\eeqa}{\end{eqnarray}}

\def\bc{\begin{center}}
\def\ec{\end{center}}

\def\to{\rightarrow}

\def\gsim{\mathrel{\mathpalette\atversim>}}

\def\bc{\begin{center}}
\def\ec{\end{center}}

\def\gsim{\mathrel{\rlap{\lower4pt\hbox{\hskip1pt$\sim$}}

    \raise1pt\hbox{$>$}}}       %greater than or approx. symbol

\def\gsim{\mathrel{\rlap{\lower4pt\hbox{\hskip1pt$\sim$}}
    \raise1pt\hbox{$>$}}}       %greater than or approx. symbol

\begin{document}
\makeatletter
\def\fmslash{\@ifnextchar[{\fmsl@sh}{\fmsl@sh[0mu]}}
\def\fmsl@sh[#1]#2{%
  \mathchoice
    {\@fmsl@sh\displaystyle{#1}{#2}}%
    {\@fmsl@sh\textstyle{#1}{#2}}%
    {\@fmsl@sh\scriptstyle{#1}{#2}}%
    {\@fmsl@sh\scriptscriptstyle{#1}{#2}}}
\def\@fmsl@sh#1#2#3{\m@th\ooalign{$\hfil#1\mkern#2/\hfil$\crcr$#1#3$}}
\makeatother
%\baselineskip 24pt

%%%%%%%%%%%%%%%%%%%%%%%%%%%%%%%%%%%%%%%%%%%%%%%%%%%%%%%%%%%%%%%%%
%%%
%%%                      TITLE PAGE
%%%
%%%%%%%%%%%%%%%%%%%%%%%%%%%%%%%%%%%%%%%%%%%%%%%%%%%%%%%%%%%%%%%%%
\thispagestyle{empty}
\begin{titlepage}
\boldmath
\begin{center}
  \Large {\bf Inflation on a Non-Commutative Space-Time}
    \end{center}
\unboldmath
\vspace{0.2cm}
\begin{center}
{  {\large Xavier Calmet}\footnote{x.calmet@sussex.ac.uk} and {\large Christopher Fritz}\footnote{c.fritz@sussex.ac.uk}}
 \end{center}
\begin{center}
{\sl Physics $\&$ Astronomy, 
University of Sussex,   Falmer, Brighton, BN1 9QH, United Kingdom 
}
\end{center}
\vspace{5cm}
\begin{abstract}
\noindent
We study inflation on a non-commutative space-time within the framework of enveloping algebra approach which allows for a consistent formulation of general relativity and of the standard model of particle physics. 
We show that within this framework, the effects of the non-commutativity of spacetime are very subtle. The dominant effect comes from contributions to the process of structure formation. We describe the bound relevant to this class of non-commutative theories and derive the tightest bound to date of the value of the non-commutative scale within this framework. Assuming that inflation took place, we get a model independent bound on the scale of space-time non-commutativity of the order of 19 TeV.
 \end{abstract}  

\end{titlepage}
%\pacs{}

%%%%%%%%%%%%%%%%%%%%%%%%%%%%%%%%%%%%%%%%%%%%%%%%%%%%%%%%%%%%%%%%
%%%
%%%                     INTRODUCTION
%%%
%%%%%%%%%%%%%%%%%%%%%%%%%%%%%%%%%%%%%%%%%%%%%%%%%%%%%%%%%%%%%%%%

\newpage
\section{Introduction}

The idea of space-time non-commutativity dates back to the early days of quantum field theory when it was hoped that it may help to make sense of UV divergences which are intrinsic to this framework \cite{Moyal:1949sk,Groenewold:1946kp}. With the advent of renormalization and the proof that physically relevant Yang-Mills theories were renormalizable, non-commutative gauge theories lost much of their appeal. However, there was a renewal of interest for such theories when they reappeared as a certain limit in string theory \cite{Seiberg:1999vs,Connes:1997cr}. In \cite{Seiberg:1999vs}, it was shown that the end points of open strings ending  on a $Dp$-brane with a Neveu-Schwarz two form flux $B$ background do not commute. String theory has an additional symmetry transformation known as T-duality, which relates geometric structures in different topologies. It naturally gives rise to non-commutative geometry. Independently of string theory, quantum gravity is likely to involve the notion of a minimal length see e.g. \cite{Calmet:2004mp,Calmet:2007vb}, which could imply a non-commutativity of space-time at short distances. This may help to alleviate the problem of the non-renormalizability of perturbative quantum gravity. 

There are different approaches to non-commutative geometry, which can be divided in roughly two classes. The first approach is due to Alain Connes. It is based on the notion of the spectral triple and has its origin in mathematical physics. The second approach indeed goes back to Moyal and Groenewold  \cite{Moyal:1949sk,Groenewold:1946kp} and emphasizes that space-time itself might be non-commutativity at short distance. The non-commutativity of space-time leads to issues with space-time and gauge symmetries.  There are two distinct ways to deal with these issues. One is to take gauge fields to be as usual Lie algebra valued and to restrict the gauge symmetries which can be considered (see e.g. \cite{Seiberg:1999vs}). The other one is to take gauge fields in the enveloping algebra  which enables one to consider any gauge group with any representation for the matter fields \cite{Madore:2000en,Jurco:2000ja,Jurco:2001rq,Calmet:2003jv,Calmet:2006zy}. In this article, we will consider the latter approach and derive the tightest bound to date on the non-commutative scale within this approach.

We shall focus here on the simplest model of space-time non-commutativity which has been extensively studied and will consider non-commuting coordinates with a canonical structure
\begin{equation}
\label{UncRel}
[\hat{x}^{\mu},\hat{x}^{\nu}] = i\theta^{\mu\nu},
\end{equation}
where $\theta^{\mu\nu}$ is a constant tensor of mass dimension -2.

Our aim is to investigate effects of space-time non-commutativity in the early universe.  We thus have to select a framework which enables us to formulate both field theories and general relativity on a non-commutative space-time. While there are different approaches to space-time non-commutativity, there is only one which leads the the well known standard model of particle physics and general relativity in the low energy regime. We shall thus use the enveloping algebra approach \cite{Madore:2000en,Jurco:2000ja,Jurco:2001rq,Calmet:2003jv,Calmet:2006zy} which enables one to formulate any gauge theory including arbitrary representations for the gauge and matter fields on a non-commutative space-time. This approach has led to a consistent formulation of the standard model of particle physics on such a space-time\cite{Calmet:2001na}. Treating General Relativity as a gauge theory, one can formulate also formulate General Relativity on a non-commutative space-time \cite{Calmet:2005qm,Calmet:2006iz,Calmet:2005mc}. It turns out that one needs to limit general coordinate transformations to those which are volume preserving diffeomorphisms. This leads to unimodular gravity which is known to be, at least classically, equivalent to general relativity. Following the enveloping algebra approach has several benefits. First of all, it makes use of real symmetries which imply a conserved charged via Noether's theorem. Such theories have an exact space-time symmetry \cite{Calmet:2004ii,Calmet:2006iy} which corresponds to Lorentz invariance in the limit of $\theta^{\mu\nu} \to 0$. The implication of this symmetry is that all the bounds on space-time non-commutativity are weak \cite{Calmet:2004dn}, typically of the order of a TeV \cite{Calmet:2006zb,Ohl:2010zf}.
 
Using this framework, we will consider inflation and the cosmic microwave background on a non-commutative space-time. There are many attempts to study inflation in the context of a non-commutative space-time  \cite{Akofor:2008gv,Akofor:2007fv,Bertolami:2002eq,Palma:2009hs,K.:2014ssa,Li:2013qga,Bertolami:2005hz,Huang:2003fw,Chu:2000ww}\footnote{These previous studies have mainly focussed on a non-commutative inflaton without considering non-commutative effects in the gravity sector. They have obtained bounds of the order of 10 TeV.}, but as far as we know this is the first study of early universe physics using the enveloping algebra approach which allows to study in details the effects of the non-commutativity of space-time on the metric. As an example we will consider chaotic inflation \cite{Linde:1983gd} on a non-commutative space-time and show that the effects of non-commutativity vanish both for the scalar field and for the metric. This is a rather surprising and interesting result since one might have expected that a preferred direction in space-time could lead to large effects in the slow role parameters since inflation could have exponentially increased the original asymmetry in space-time. We then consider the effects of space-time non-commutativity on the CMB which are this time non vanishing. This is not surprising as non-commutative gauge theories are a special case of non-local theories which are known to affect the CMB. We derive the tightest bound to date on the scale of space-time non-commutativity within this framework.

\section{Theoretical framework}

We consider here the algebra $\hat{\mathcal{A}}$ of non-commutative space-time coordinates $\{ \hat{x}^{\mu}\}$ which satisfying the canonical relation
\begin{equation}
\label{ComRel}
[\hat{x}^{\mu},\hat{x}^{\nu}] = i\theta^{\mu\nu},
\end{equation}
where $\theta \in \Omega^{2}(T \mathcal{M})$ is a constant tensor and can be locally expressed as $\theta = \theta^{\mu\nu} \partial_{\mu} \otimes \partial_{\nu}$ with $\theta^{\mu\nu} = -\theta^{\nu\mu}$. As usual, we want to represent functions in $\hat{\mathcal{A}}$ as elements in the space of linear complex functions $\mathcal{F}$. To do so we introduce the Moyal star product
\begin{equation}
\label{Moyal}
\renewcommand{\arraystretch}{1.5}
\begin{array}{ll}
(f_{1} \cdot f_{2})(\hat{x}) & = (f_{1} \star f_{2})(x) =  \displaystyle \sum^{\infty}_{n=0} \left( \frac{i}{2} \right)^{n}\frac{1}{n!} \theta^{\mu_{1}\nu_{1}} \cdots \theta^{\mu_{n}\nu_{n}} \partial_{\mu_{1}}  \cdots \partial_{\mu_{n}} f_{1} \partial_{\nu_{1}} \cdots \partial_{\nu_{n}} f_{2}.
\end{array}
\end{equation}
Before continuing on to the main discussion, it will be useful to note some useful properties of the star product. Firstly, under complex conjugation one has
\begin{equation}
\label{ComplexConj}
(f_{1} \star f_{2})^{\ast} = f_{2}^{\ast} \star f_{1}^{\ast}.
\end{equation}
Secondly, the trace property under integration implies that
\begin{equation}
\label{PartialInt}
\int d^{4}x  (f_{1} \star f_{2})(x) = \int d^{4}x  (f_{1} \cdot f_{2})(x)
\end{equation}
and more generally, one also has the cyclicity property
\begin{equation}
\label{CycInt}
\renewcommand{\arraystretch}{1.5}
\begin{array}{ll}
\displaystyle \int d^{4}x  (f_{1} \star \cdots \star f_{n})(x) 
& = \displaystyle \int d^{4}x  (f_{1} \star \cdots \star f_{m-1}) \cdot (f_{m} \star \cdots \star f_{n})(x) \\
& = \displaystyle \int d^{4}x  (f_{m} \star \cdots \star f_{n}) \cdot (f_{1} \star \cdots \star f_{m-1})(x).
\end{array}
\end{equation}

It is important to note, given that $\theta$ is constant, that this theory violates general diffeomorphism invariance. However, as shown in \cite{Calmet:2005qm} we may recover a reduced group of diffeomorphisms compatible with \eqref{ComRel} parametrized by 
\begin{equation}
\label{DiffeoInv}
\hat{x}'^{\mu}  = \hat{x}^{\mu} + \hat{\xi}^{\mu}.
\end{equation}
A subset of these transformations given by
\begin{equation}
\label{VolDiffeoInv}
\hat{\xi}^{\mu} = \theta^{\mu\nu} \partial_{\nu} \hat{f}(\hat{x})
\end{equation}
leave $[\hat{x}^{\mu},\hat{x}^{\nu}] = i\theta^{\mu\nu}$ invariant. We shall thus only consider such transformations.  Note that the Jacobian of these transformation is equal to one. The transformations which preserve the non-commutative algebra correspond  to the reduced group of diffeomorphisms which are volume preserving. In other words, on a non-commutative space-time, we are forced to consider unimodular gravity. This is the main difference between our work and precious attempts at formulating inflation on a non-commutative space-time\cite{Bertolami:2002eq,Fang:2007ba,Palma:2009hs,Li:2013qga}. The approach to general relativity on a non-commutative space-time formulated in  \cite{Calmet:2005qm}  relies on gauging a local SO(3,1) (the tetrad approach). The local SO(3,1) gauge symmetry is implemented using the enveloping algebra approach. This means that the gauge fields are assumed to be in the enveloping algebra instead of the usual Lie algebra. The local gauge invariance in enforced using the Seiberg-Witten maps order by order in $\theta$ \cite{Calmet:2005qm}. We now have all the tools needed to formulate a consistent scalar field action in a curved space-time on a non-commutative space-time. 

\section{Non-Commutative Scalar Action}

We consider inflation driven a single scalar field with a potential $V(\phi^{n})$ and denote for convenience $\phi \equiv \phi(x)$. In the commutative case, the action may be written
\begin{equation}
\label{CInflaton}
S=\int d^{4}x e \left( \frac{1}{2} \partial_{\mu} \phi \partial^{\mu} \phi - \frac{1}{2}m^{2} \phi^{2} - \sum_n  c_n \frac{\phi^{n}}{\Lambda^{(n-4)}} \right),
\end{equation}
where $e$ is the tetrad determinant, $\Lambda$ in an energy scale and $c_n$ are dimensionless Wilson coefficients of order unity.  The choice of this frame follows from the derivation of non-commutative general relativity from the Seiberg-Witten map, as in \cite{Calmet:2005qm,Calmet:2006iz}, for which gravity is treated as a gauge theory. Another reason is that when mapping quantities on to a non-commutative space, it is very difficult to do so for a square root (which may not even exist in $\mathcal{A}_{\theta}$) and $e$ is used as an effective way to represent $\sqrt{g}$. Setting the tetrad determinant to one, the action for the non-commutative scalar field may be written
\begin{equation}
\label{NCInflaton}
S=\int d^{4}x \left( \frac{1}{2}G^{\mu\nu} \star \partial_{\mu} \phi \star \partial_{\nu} \phi - \frac{1}{2}m^{2}\phi \star \phi  - \sum_n c_n \frac{\phi^{n\star}}{\Lambda^{(n-4)}} \right).
\end{equation}

One might be tempted to take $\partial_{\mu} \phi \star \partial^{\mu} \phi$ and use \eqref{PartialInt} to eliminate the star product, as is done with, e.g., the mass term. However, this is not possible here because of the space-time dependent metric. Let us add a quick comment on our conventions here: when defining a derivative operator in $\mathcal{A}_{\theta}$, one naturally has a map $\partial^{\star}_{\mu}:\mathcal{A}_{\theta} \rightarrow \mathcal{A}_{\theta}$ whose (left) action is defined to be $\partial^{\star}_{\mu} \rhd f \equiv \partial_{\mu} f$ with $f \in \mathcal{A}_{\theta}$. However, the same definition does not hold for $\partial^{\star\mu}$, which, in general, is a power series in $\theta$ and a higher order differential operator. It may be obtained from the relation $\partial^{\star\mu} = G^{\mu\nu} \star \partial_{\nu}$. For the non-commutative metric, the condition $G^{\mu\nu}|_{\theta=0} = g^{\mu\nu}$ applies. Furthermore, we are free to choose a frame where $G^{\mu\nu} = g^{\mu\nu}$ but we must keep in mind that $G^{\mu\nu} \star G_{\mu\alpha} \neq \delta^{\mu}_{\ \alpha}$.  We thus require a `star inverse' to be defined such that $G^{\mu\nu} \star G_{\star\mu\alpha} = \delta^{\mu}_{\ \alpha}$, see for example \cite{Aschieri:2005yw}. It will however, not be necessary for the analysis presented here.  Indeed making this choice for the metric leads to a significant simplification.  It is unnecessary to find an expansion of $G^{\mu\nu}$ in terms of $\theta$. 

We now expand out the star products in the action, mapping the non-commutative theory to a commutative space-time. Note that, as we have just explained, the Seiberg-Witten map for the metric is trivial. For the kinetic term we find
\begin{equation}
\label{new4}
g^{\mu\nu} \star \partial_{\mu} \phi \star \partial_{\nu} \phi =  g^{\mu\nu} \partial_{\mu} \phi \partial_{\nu} \phi -\frac{1}{8} g^{\mu\nu} \theta^{\alpha\beta} \theta^{\gamma\delta} \partial_{\alpha} \partial_{\gamma} \partial_{\mu} \phi \partial_{\beta} \partial_{\delta} \partial_{\nu}  \phi+ \mathcal{O}(\theta^{4})
\end{equation}
 and for the potential we get
\begin{equation}
\label{ReScMoyExp}
\renewcommand{\arraystretch}{1.5}
\begin{array}{ll}
\phi^{n\star}
& = \displaystyle \phi^{n} - \frac{1}{8} \theta^{\mu\nu}\theta^{\alpha\beta} \sum^{n-2}_{m=1} \phi^{n-m-1} \partial_{\nu} \partial_{\beta} \phi^{m} \partial_{\mu} \partial_{\alpha} \phi + \mathcal{O}(\theta^{3}),
\end{array}
\end{equation}
where \eqref{CycInt} has been used. We see non-commutative corrections appear only at second order in $\theta$ meaning that any effects are going to be strongly suppressed. It is worth noting the appearance of corrections in the kinetic term. This feature is absent in \cite{Bertolami:2002eq,Fang:2007ba,Li:2013qga}. In a flat space-time, one may use the cyclicity of the star product to cancel corrections to quadratic terms such as these. However, since we are dealing with a curved space-time, we cannot do this here. 

We now have all the tools to consider inflation on a non-commutative space-time using the enveloping algebra approach. Here is the set of assumptions we are making. Firstly, the inflaton field is taken to be homogenous i.e. $\phi \equiv \phi(t)$. Secondly, we assume that the same is true of the metric for a spatially flat, FLRW like cosmology: We know that there are second order in $\theta$ corrections to Einstein's equations as shown in \cite{Calmet:2005qm}, but these corrections vanish for a metric which is purely time-dependent. Feeding these assumptions into the above equations, one quickly sees that, owing to the antisymmetry of $\theta^{\mu\nu}$, all but the zeroth order terms vanish. We thus see that the inflation is does not feel the non-commutativity of space-time. In particular the slow roll parameters are given by their usual commutative expression
\begin{equation}
\label{SlowRollParamSt}
\renewcommand{\arraystretch}{1.5}
\begin{array}{cc}
\displaystyle \epsilon = \displaystyle \frac{M_{P}^{2}}{2} \left(\frac{1}{V_{0}}\frac{\partial V_{0}}{\partial \phi}\right)^{2} &
\displaystyle \eta = \displaystyle \frac{M_{P}^{2}}{V_{0}} \left|\frac{\partial^{2} V_{0}}{\partial \phi^{2}}\right|. 
\end{array}
\end{equation}
This result is somewhat surprising; intuitively, one would expect that the presence of a preferred direction in space-time would result in anisotropic contributions to the metric at some order in $\theta$. However, the nature of the corrections is such that they vanish to all orders conserving the initial isotropy. It is interesting to consider this against cosmological paradigms, such as the flatness problem, which are generally amplified throughout time. As usual, space-time non-commutative effects are very elusive \cite{Calmet:2004dn}!

\section{CMB Corrections}

While, within our framework, there are no effects of space-time non-commutativity on the slow role parameters, we now show that there are interesting observable effects on the CMB.  A homogenous field may not have any corrections, but the same is not necessarily true of perturbations to that field.
 \begin{equation}
\label{InflatonPert}
\phi(t,\mathbf{x}) = \phi(t) + \delta \phi(t,\mathbf{x}).
\end{equation}
While the overall evolution of the universe may be unaffected, space-time non-commutativity could have some influence on structure formation. We thus need to consider non-commutative corrections to inflaton perturbations. It is well known that general relativity and unimodular gravity, at least in the classical regime, are equivalent \cite{Unruh:1988in}. It is generally possible to find a subset of spacetime where we can write Einstein's equations such that $det(g_{\mu\nu})=1$. This implies that the predictions for inflation in unimodular gravity are the same as in the full general relativity framework on a classical space-time. This has been explicitly shown in \cite{Cho:2014taa}. As emphasized already, our approach to general relativity formulated on a non-commutative space-time forces us to consider unimodular gravity. However, the work in \cite{Cho:2014taa} implies that the details of the analysis of small perturbations does not depend on whether the underlying theory of gravity is general relativity or unimodular gravity. The analysis performed in \cite{Akofor:2008gv,Akofor:2007fv,K.:2014ssa} where statistical anisotropies of the CMB were studied without paying attention to non-commutative corrections to general relativity thus applies to the enveloping algebra.  However, our framework enables us to justify the assumption that non-commutative corrections to metric can be neglected.  Indeed, using the results in presented in \cite{Calmet:2005qm,Calmet:2006iz}, it is straightforward to see that for a purely time-dependent metric such as the FLRW metric, the non-commutative corrections to the classical metric vanish to all orders in $\theta$. 

The calculation of the $n$-point correlators takes place at the level of the equations of motion. This calculation will be unaltered for a unimodular metric and will thus apply to the enveloping algebra approach considered here. We can thus follow the technique developed in \cite{Akofor:2008gv,Akofor:2007fv,K.:2014ssa}. We first consider the  contributions from non-commutativity to the power spectrum of the CMB which are obtained from calculating the two-point correlation function for scalar perturbations. For a co-moving (commutative) scalar field $\tilde{\zeta} (\eta,\mathbf{k})$, where the tilde indicates that this is a Fourier mode of $\zeta (\eta,\mathbf{x})$, the power spectrum in terms of the two-point function is  
\begin{equation}
\label{PSpecDef}
\langle 0 | \tilde{\zeta}^{\dagger} (\eta,\mathbf{k}) \tilde{\zeta} (\eta,\mathbf{k'})|0 \rangle = (2\pi)^{3} P_{\theta=0} (\eta,\mathbf{k}) \delta^{3}(\mathbf{k} - \mathbf{k'}).
\end{equation}
We have defined the conformal time
 \begin{equation}
\label{conftime}
d\eta = \frac{1}{a(t)} dt,
\end{equation}
where $a(t)$ is the cosmological scale factor. At the time horizon crossing $\eta_{0}$, this is given by
\begin{equation}
\label{PSpecCom}
 P_{0} (\eta,\mathbf{k}) = \frac{16\pi}{9E} \left. \frac{H^{2}}{2k^{3}} \right|_{a(\eta_{0})H=k}.
\end{equation}
It was found in \cite{Akofor:2007fv}, that this is modified by non-commutativity and that a revised expression for the power spectrum can be derived.
\begin{equation}
\label{PSpecNC}
 P_{\theta} (\eta,\mathbf{k}) = P_{0} (\eta,\mathbf{k}) \cosh(H \theta^{0a} k_{a}),
\end{equation}
where $H$ is the Hubble parameter. Expanding to leading order gives

\begin{equation}
\label{PSpecNCExp}
 P_{\theta} (\eta,\mathbf{k}) = P_{0} (\eta,\mathbf{k}) + \frac{H^{2}}{2!} P_{0} (\eta,\mathbf{k}) \theta^{0a} \theta^{0b} k_{b} k_{a}.
\end{equation}
Again, we note the absence of first order contributions in the non-commutative parameter. Also interesting to note is that a more general treatment of rotational invariance violation in \cite{Ackerman:2007nb} gives a similar result. We now see that non-commutativity indeed has an effect that may be measured by CMB experiments.

By doing so, bounds on the scale of space-time noncommutativity have been derived in \cite{Akofor:2007fv} using WMAP, ACBAR and CBI and in \cite{K.:2014ssa} using PLANCK data. In \cite{K.:2014ssa} the authors found a bound of 19 TeV on the scale of spacetime non-commutativity. Since the same derivation goes through in our formalism as well, this leads to the tightest bound to date on the energy scale of spacetime non-commutativity within the framework of the enveloping algebra approach.
 
\section{Conclusion}

We have considered corrections induced by non-commutativity on a scalar inflaton field. Specifically, we consider the enveloping algebra approach to space-time non-commutativity with a constant non-commutative parameter. In this approach the reduced group of diffeomorphisms, chosen so as to leave \eqref{ComRel} invariant, leads to  unimodular gravity. By replacing conventional multiplication by the Moyal star product and expanding in terms of the non-commutative parameter $\theta$ as well as using the Seiberg-Witten maps for the local SO(3,1) gravitational theory, it was shown that no corrections enter the inflationary action. This leads us to the somewhat surprising realization that even in the presence of a preferred direction in space-time, does not affect the overall evolution of the universe. Instead, one must examine primordial perturbations to the inflaton field which do experience the non-commutativity and look for their imprints in the CMB. Owing to the classical equivalence of general relativity and unimodular gravity, the analysis necessary for doing so is the same in both cases. We can derive the bound $\sqrt{\theta}$ to $\sim19 \text{ TeV}$ within our approach to space-time non-commutativity. This is the tightest limit to date on the scale of space-time non-commutativity within the enveloping algebra approach to space-time non-commutativity. 

{\it Acknowledgments:}
This work is supported in part  by the Science and Technology Facilities Council (grant number  ST/J000477/1).

%%%%%%%%%%%%%%%%%%%%%%%%%%%%%%%%%%%%%%%%%%%%%%%%%%%%%%%%%%%%%%%%%
%%%
%%%                     BIBLIOGRAPHY
%%%
%%%%%%%%%%%%%%%%%%%%%%%%%%%%%%%%%%%%%%%%%%%%%%%%%%%%%%%%%%%%%%%%%

\bigskip{}

\end{document}